# "Mathematical and Preclinical Investigation of Respiratory Sinus Arrhythmia Effects on Cardiac Output"

Sahar Rahbar

Sahar.rahbar@mail.um.ac.ir

**Abstract**

Respiratory sinus arrhythmia (RSA) is heart rate variability in synchrony with respiration although its functional significance not clear. The loss of sinus arrhythmia may indicate underlying heart failure or disease; therefore, there would be a great advantage of knowing how it works and affects the cardio-respiratory system, especially by providing a mathematical model. To this end, Windkessel model and cardiovascular partial differential equations are used to obtain cardiac output based on the elasticity of left ventricle, which is related to RSA. By solving the corresponding equations, it would be possible to propose a new model to predict the RSA effects on cardiac output.

**Introduction**

Studies related to respiration and heart rate variation have been a great controversy over a century and have continued as a field of study to the present day. The acceleration of the heart rate after inspiration and deceleration of it during expiration, respiratory sinus arrhythmia (RSA) [1], which is used as an index of cardiac vagal tone and contributes to heart rate variability [2]. The R-R interval on an ECG is an indication of RSA which is shortened and prolonged during inspiration and expiration correspondingly [3]. Although the mechanism of RSA and the factors affecting it have been studied extensively, the physiological significance of RSA remains elusive [2]. To explain RSA phenomena, researches have focused on three primary mechanisms as follow, a) There is a direct communication between the respiratory center and the heart-rate centers by central nervous system [4]. Heart rate is controlled by the activity of premotor cardio-inhibitory parasympathetic neurons (CPNs) in the brainstem, and RSA is mediated in part by central respiratory modulation of CPN activity [5], b) An indirect control through blood pressure, c) A reflex control in response to stretch receptors in the lungs and chest [4]. It is widely accepted that the loss of RSA is a prognostic indicator for cardiovascular disease and the prominent presence of RSA indicates a healthy cardiac system; however, the reasons for this are still being debated [6].

Hayano et al. (1996) suggested that matching ventilation and perfusion in the lungs is the physiological function of RSA [7] which in turn leads to optimize the amount of oxygen uptake and carbon dioxide removal. Moreover, Hayano and Yasuma (2003) hypothesized that RSA is an intrinsic resting function of the cardiopulmonary system and decreases the energy expenditure of the cardiopulmonary system by reducing the number of heart beats during expiration [8]. Furthermore, Giardino et al. (2003) by measuring the equivalents of carbon dioxide and oxygen during paced breathing claimed that gas exchange efficiency increases with RSA [9].

The process of gas exchange in the lungs has been modeled in 2006 and studied from different perspectives in detail at different levels. Ben-Tal (2006) concerned these interactions via the gas exchange process which is regarded as the controlled system. He consumed the pleural pressure and the cardiac output as the controlled variables and the partial pressures of oxygen and carbon dioxide as the measured variables [10]. In 2012, Ben-Tal et al. used the model of gas exchange in mammals to calculate gas exchange efficiency during fast, normal and slow paced breathing. They found that although gas exchange efficiency improved with slow and deep breathing and also with increased mean heart rate (HR), this was unrelated to RSA. However, the model used in their study did not include feedback mechanisms of the cardio-respiratory system and numerical simulations were performed by pre-setting the heart rate variations [2]. He also used the model with direct central respiratory modulation of the parasympathetic cardiac signal as the main mechanism for RSA and includes blood pressure control via the baroreflex in 2014. They confirmed that RSA minimizes the work done by the heart while maintaining arterial carbon dioxide [6].

In order to continue research in this area, a simple quantitative model of cardiovascular variability was proposed by Buchner in 2018 that presented the relationship between the RSA and blood pressure fluctuations [11]. O'Callaghan et al. (2019) reported that reinstatement of RSA would improve cardiac function in rats with heart failure by ventricular remodeling. They proposed that RSA pacing reverse-remodels the heart in heart failure, thus it could be considered as a new form of cardiac pace-making. In other words, they suggest that RSA optimize oxygen delivery to the heart with increasing workload [1]. Moreover, O'Callaghan et al. noted that respiratory modulated heart rate (RMH) pacing did not improve the ejection fraction; nevertheless, Samuel H.H. Chan (2020) suggested the main reason could be that cardiac pacing is directed primarily against reduced heart rate. He claims that a combination of RMH pacing and cardiac contractility modulation (CMM) may be better for patients who suffer from decrease in heart rate and ejection fraction [12].

Indeed, RSA could be an indicator for diagnosing cardiovascular disease and sudden cardiac death. Therefore, during reduced coronary blood flow and oxygen supply, the absence of RSA may promote cardiac remodeling [13]. Many questions are still existed in cardiovascular regulation and providing related answers may not only push forward the studies, but also provide new methods to diagnose the risks and lead to more efficient patient handling. As Maja Elstad et al. (2018) stated, providing mathematical and computational models are needed to better understanding the putative role that RSA plays in stabilizing systemic blood flow at respiratory frequencies and also to examine and test hypotheses not easily tested in experimental models without ethical concern [13].

Considering strong evidence that a loss of RSA in cardiorespiratory diseases would affect cardiac work efficiency, has led to the idea to investigate a mathematical modeling of the effect of RSA on cardiac output and also survey the results on large animals in order to verify the outcomes. Interaction of RSA on cardiac output is not fully understood and provides motivation for developing the mathematical models presented in the following section. The preclinical part could be carried out by using analogue biofeedback device and modulating cardiac pacemaker spike frequencies as stated in [1]. The elasticity commonly exhibits a longer term variation which is one of the main causes of RSA [14]. The purpose of this study is to present a mathematical modeling of the cardiac output and RSA, which is dependent on elasticity of left ventricle.

**Methods**

Our study consists of four sections: model development, numerical and analytical solution, preclinical analysis and verifying the investigation. To present the final model, first the cardiovascular partial differential equations and Windkessel model will be described.

1- Mathematical model: The following differential equations for cardiovascular system in order to assessing the blood pressure curve are presented below [14].

$$C_e \dot{P}_e = \frac{P_{LV} - P_e}{R_e(t)} - \frac{P_e - P_a}{R_a} + f(t)$$

$$C_a \dot{P}_a = \frac{P_e - P_a}{R_a} - \frac{P_a - P_v}{R_v}$$

$$C_v \dot{P}_v = \frac{P_a - P_v}{R_v} - \frac{P_v - P_{LV}}{R_{LV}}$$

$$\frac{d}{dt}\left(\frac{P_{LV}}{E_{LV}(t)}\right) = \frac{P_v - P_{LV}}{R_{LV}} - \frac{P_{LV} - P_e}{R_e(t)} - f(t)$$

These equations reproduce the dominant features of the blood pressure curve, the dicrotic notch (a small dip in the blood pressure curve associated with aortic valve closure), and the variation due RSA. In all equations, P is pressure, C is compliance coefficient and R can be thought as the effective resistance to flow. The subscripts e, a, v, LV represent exit, arterial, venous and left ventricle and dots indicate the derivative with respect to time. Also, $f(t) = c_4 \exp(-\frac{c_5(t-t_n-c_6\Delta t)^2}{(\Delta t)^2})$ presents a model of the dicrotic notch. The constants $c_4$, $c_5$, $c_6$ indicate the height of the pulse, the sharpness and the position of the center. The resistances of aortic valve must be time dependent as follow:

$$R_e(t) = R_{e0}\left[1 + \epsilon_1(\exp(-A_1(p_{LV} - p_e))\right]$$

where $R_{e0}$ is the constant value when the valve is fully open. Also, we consider $\epsilon_1 = 10^{-5}$ and $A_1 = 0.5$.

The elasticity of left ventricle is presented as follow:

$$E_{LV} = E_d + a(t)(E_s - E_d)$$

$$a(t) = \begin{cases} \sin^2\left(\dfrac{3\omega(t-t_0)}{2}\right) & t - t_0 \in (0, T_{sys}) \\ 0 & t - t_0 \in (T_{sys}, T) \end{cases}$$

where the diastolic elasticity $E_d$ is constant. Moreover, to model RSA with heart rate varying with respiration, $\omega$ would be:

$$\omega = \omega_0 + c_3 \sin\left(\dfrac{\omega_0 t}{c_2}\right)$$

The peak in the elasticity height varies over the longer time-scale associated with RSA. Considering this, the systolic elasticity would be:

$$E_s = E_{s0} + c_1 \sin\left(\dfrac{\omega_0 t}{c_2}\right)$$

where the constants $c_1, c_2, c_3$ represent half the variation of elasticity height, the number of heart beats per respiration (typically around 5) and half the variation of $\omega$, respectively. Typical parameter values are presented in table 1.

*Table 1: Parameter values of cardiovascular equations* [14]

| Parameter | Value | Units | Parameter | Value | Units |
|---|---|---|---|---|---|
| $c_1$ | 0.1 | mmHg/ml | $c_2$ | 6 | Beats/breath |
| $c_3$ | 0.01 | $s^{-1}$ | $c_4$ | 500 | |
| $c_5$ | 4log100 | | $c_6$ | 7.5 | |
| $R_a$ | 0.06 | s.mmHg/ml | $R_v$ | 0.016 | s.mmHg/ml |
| $R_{LV}$ | 1.2 | s.mmHg/ml | $R_{e0}$ | 0.016 | s.mmHg/ml |
| $C_e$ | 1.5 | ml/mmHg | $C_a$ | 1.5 | ml/mmHg |
| $C_v$ | 50 | ml/mmHg | $T$ | 0.9 | s |
| $T_{sys}$ | T/3 | s | $E_d$ | 0.06 | mmHg/ml |
| $E_{s0}$ | 3 | mmHg/ml | $\omega_0$ | 7.54 | |

2- Windkessel model: Cardiac output (CO) is a key hemodynamic variable that is commonly used to establish differential diagnoses and monitor disease progression. Figure 1 shows a Windkessel model which describes the basic morphology of an arterial pressure pulse [15].

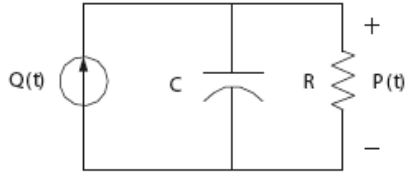

*Figure 1. Circuit representation for the Windkessel model.*

In the electrical circuit analog in Figure 1, a single resistor R, representing total peripheral resistance (TPR), and a single capacitor C, representing the aggregate elastic properties of all systemic arteries and the pumping action of the heart is presented by Q(t). By using Windkessel model, the cardiac output can be estimated from the following equation [15].

$$CO_n = C_n(\frac{\Delta P_n}{T_n} + \frac{\bar{P}_n}{\tau_n})$$

where $\Delta P_n$ is the beat-to-beat pressure change at the onset times and $\bar{P}_n$ is the average arterial blood pressure over the $n^{th}$ cycle. Moreover, $T_n$ is the duration of the $n^{th}$ cardiac cycle, i.e., the beat that begins at time $t_n$ and ends at time $t_{n+1}$ (so $T_n = t_{n+1} - t_n$) and $\tau_n = C_n R_n$, during which $\Delta P_n$ and $\bar{P}_n$ can be calculated as

$$\Delta P_n = P(t_{n+1}) - P(t_n)$$

$$\bar{P}_n = \frac{1}{T_n}\int_{t_n}^{t_{n+1}} P(t)dt$$

This proposal aims to connect the pressure in CO equation to the pressure and elasticity of left ventricle which lead to a non-linear system of differential equations. By solving these equations and using the cardiac output equation, the effect of RSA on CO in the case of RSA presence and absence can be compared. Also, preclinical study would provide data to verify the model. This model would be relatively simple compared to some of the previously-published models. However, it can still reproduce a wide range of physiological observations and it can be used to study the causes and benefits of RSA, as well as making novel predictions.

**Limitation of study**

Challenges in solving a system of non-linear differential equations and also making a precise approximation by a linear system would make some limitation thorough the study. Furthermore, obtaining the relevant data from the animals may bring about some challenges that maybe uncover in the process of handling the research/recording and measuring the desired data.

**References**

[1] Erin L O'Callaghan, Renata M Lataro, Eva L Roloff, Ashok S Chauhan, Helio C Salgado, Edward Duncan, Alain Nogaret, Julian F R Paton, "Enhancing respiratory sinus arrhythmia increases cardiac output in rats with left ventricular dysfunction," *The journal of physiology,* 2019.

[2] A. Ben-Tal, S. S. Shamailov and J. F. R. Paton, "Evaluating the physiological significance of respiratory sinus arrhythmia: looking beyond ventilation–perfusion efficiency," *The Journal of Physiology,* p. 1989–2008, 2012.

[3] Fumihiko Yasuma and Jun-ichiro Hayano, "Respiratory Sinus Arrhythmia- Why Does the Heartbeat Synchronize With Respiratory Rhythm?," *chestjournal,* vol. 125, no. 2, pp. 683-690, 2004.

[4] B. F. WOMACK, "The Analysis of Respiratory Sinus Arrhythmia Using Spectral Analysis and Digital Filtering," *IEEE TRANSACTIONS ON BIO-MEDICAL ENGINEERING,* Vols. BME-18, no. 6, pp. 399-409, 1971.

[5] Robert A. Neff, Jijiang Wang, Sunit Baxi, Cory Evans, David Mendelowitz, "Respiratory Sinus Arrhythmia-Endogenous Activation of Nicotinic Receptors Mediates Respiratory Modulation of Brainstem Cardioinhibitory Parasympathetic Neurons," *Circulation Research,* pp. 565-572, 2003.

[6] Alona Ben-Tal , Sophie S. Shamailov , Julian F.R. Paton , "Central regulation of heart rate and the appearance of respiratory sinus arrhythmia: New insights from mathematical modeling," *Mathematical Biosciences,* vol. 255, p. 71–82, 2014.

[7] Jouasset-Strieder D, Cahill JM, Byrne JJ, "Pulmonary capillary blood volume in dogs during shock and after retransfusion," *J Appl Physiol.,* vol. 21, p. 365–369, 1966.

[8] Hayano J1, Yasuma F., "Hypothesis: respiratory sinus arrhythmia is an intrinsic resting function of cardiopulmonary system," *Cardiovasc Res,* vol. 58, p. 1–9, 2003.

[9] Giardino ND, Glenny RW, Borson S, Chan L., "Respiratory sinus arrhythmia is associated with efficiency of pulmonary gas exchange in healthy humans," *Am J Physiol Heart Circ Physiol,* vol. 284, p. H1585–H1591, 2003.

[10] A. Ben-Tal, "Simplified models for gas exchange in the human lungs," *Journal of Theoretical Biology,* vol. 238, p. 474–495, 2006.

[11] T. Buchner, "A quantitative model of relation between respiratory-related blood pressure fluctuations and the respiratory sinus arrhythmia," *Medical & Biological Engineering & Computing,* vol. 57, p. 1069–1078, 2019.

[12] S. H. Chan, "Reinstatement of respiratory sinus arrhythmia as a therapeutic target of cardiac pacing for the management of heart failure," *Journal of Physiology,* 2020.

[13] Maja Elstad, 1 Erin L. O'Callaghan, Alex J. Smith,Alona Ben-Tal and Rohit Ramchandra, "Cardiorespiratory interactions in humans and animals: rhythms for life," *Integrative Cardiovascular Physiology and Pathophysiology,* vol. 315, p. H6–H17, 2018.

[14] T. G. Myers , Vicent Ribas Ripoll, Anna Sáez de Tejada Cuenca, Sarah L. Mitchell and Mark J. McGuinness, "Modelling the cardiovascular system for assessing the blood pressure curve," *Mathematics-in-Industry Case Studies,* vol. 8, no. 2, pp. 1-16, 2017.

[15] TA Parlikar, T Heldt, GV Ranade, "Model-Based Estimation of Cardiac Output and Total Peripheral Resistance," *Computers in Cardiology ,* vol. 34, p. 379–382, 2007.